\documentclass[%
 reprint,
 amsmath,amssymb,
 aps,
]{revtex4-2}

\usepackage{graphicx}
\usepackage{dcolumn}
\usepackage{bm}
\usepackage{dutchcal}
\usepackage{lipsum, babel}

\begin{document}

\preprint{APS/123-QED}

\title{Nearly cosine series and generalized trigonometric functions}

\author{A. Curcio}
\affiliation{Sapienza University, SBAI Department,Via Scarpa, 14, 00161, Rome, Italy}

\author{G. Dattoli}
\author{E. Di Palma}
\affiliation{C.R. Enea Frascati, Nuclear Department, Via E. Fermi, 45, 00044, Frascati (Rome), Italy}
\author{P. Natalini}
\affiliation{Dipartimento di Matematica e Fisica, Università degli Studi Roma Tre, Largo San Leonardo Murialdo, 1, 00146, Rome, Italy}
\author{P. E. Ricci}
\affiliation{Dipartimento di Matematica, International Telematic University UniNettuno, Corso Vittorio Emanuele II, 39, 00186, Rome, Italy}





\date{\today}

\begin{abstract}
A class of overlooked trigonometric-like functions is explored in this article, along with the relevant applications in applications. We show indeed that Taylor series, resembling that of an ordinary cosine, are  representative of wider classes of functions, naturally suited for prolems ranging from molecular to Laser Physics. The article goes through the original motivations of the proposal and studies the relevant properties within the context of an Umbral interpretation. Their use in applications is discussed within the framework of Free Electron Laser theory, Lennard- Jones potentials and Kramers-Kronig causality identities.
\end{abstract}

\maketitle

\onecolumngrid
\section{Introduction}

\label{sec1}
The possibility of defining functions which are an extension, with non-trivial properties, of “popular” elementary forms, like those belonging to the trigonometric family is an interesting and far reaching opportunity. Which eventually leading to useful forms playing a non-secondary role either in pure and applied math.
This perspective had been pursued since the time of Euler \cite{sandifer2006euler}, who asked himself the question whether Taylor expansions of the type \cite{euler1795radicibus}

\begin{equation}
\label{GrindEQ__1_}
\begin{array}{l}
{\cos _{m} (x)=\sum _{r=0}^{\infty }\dfrac{(-1)^{r} }{\left(m+1\right)_{2r} }  x^{2r} } \\ \\
{(d)_{r} =\dfrac{\Gamma (d+r)}{\Gamma (d)} \equiv {\rm Pochhammer\; symbol}}
\end{array}
\end{equation}

yields a useful generalization, eventually leading to new classes of functions. After more than two centuries, we know that  the previous series can be framed within the context of hyper-geometric functions \cite{andrews1998special} and are recognized to open wider scenarios in the study of elementary transcendent and not only. It was Euler himself who pointed out that, for some positive integer values of $m$ (see below), the series \eqref{GrindEQ__1_} (nowadays called nearly cosine) reduces to combinations of elementary functions.

The question raised by Euler was motivated by the search for an extension of the technique leading to the solution of the Basel problem \cite{sangwin2001infinite}, regarding the sum of the infinite series $\sum _{n=1}^{\infty }\frac{1}{n^{2} } $. Euler had obtained the right answer to a problem lasting more than a century, through a clever use of the roots associated with the equation $\cos (x)=0$.

Even though no further positive accomplishment had been obtained within this respect, his suggestion, viewed from a modern point of view, has opened interesting perspectives regarding the possibility of their applications in a wide range of physical topics.

In recent times the study of new forms of trigonometric functions \cite{dattoli2017circular} has opened wider scenarios on the theory of special functions, allowing a natural transition between elementary and higher order transcendental functions.

In this article we emphasize that the original proposal can be straightforwardly translated within the point of view of the umbral calculus \cite{roman1978umbral, licciardi2022guide}. The umbral calculus has gone through an evolution from the ``classical'' perspective of Roman and Rota \cite{roman1978umbral} to the formulation summarized in ref. \cite{licciardi2022guide}, in which the use of the concept of the umbral image (UI) \cite{dattoli2023umbral} plays a central role to simplify the theory of higher order transcendental functions.

We use here the last conception, applied to generalized forms of trigonometric functions, whose UI is provided by Lorentzian functions.

The article consists of three further sections, including the consequences of the UI interpretation of functions of the type \eqref{GrindEQ__1_}, the comparison with previous researches and contains comments, regarding the relevance of the present analysis to an alternative point of view to old and new problems in the theory of special functions. We eventually conclude with an analysis of the framing of this family of functions within the realm of Bessel type series.

Our motivation goes beyond the interest of framing these functions within the context of umbral calculus, but for their genuine importance in applications. We point out indeed that the they naturally appears in the formulation of the Free Electron Laser (FEL) theory (see Sec. \ref{sss4}). We furthermore prove how, a deeper study of their underlying mathematical structure, provides significant simplifications of the computations of the physical quantities associated with FEL type devices.

\section{Umbral Images of Euler trigonometric functions}

We proceed as outlined in refs. \cite{licciardi2022guide,dattoli2023umbral} and introduce the umbral operator $\hat{\chi }$, defined in such a way that, acting on the corresponding vacuum $\phi _{0}$, yields

\begin{equation}
\label{GrindEQ__2_}
\begin{array}{l}
{{}_{m} \hat{\chi }^{r} \phi _{0} =\dfrac{1}{(m+1)_{r} } } \\ \\
{r\in R} \nonumber
\end{array}
\end{equation}
Taking also note that

\begin{equation}
\label{GrindEQ__3_}
{}_{m} \hat{\chi }^{s} {}_{m} \hat{\chi }^{r} \phi _{0} =\dfrac{1}{(m+1)_{r+s} }
\end{equation}
We can write the series in equation \eqref{GrindEQ__1_} as

\begin{equation}
\label{GrindEQ__4_}
\cos _{m} (x)=\sum _{r=0}^{\infty }(-1)^{r}  \left({}_{m} \hat{\chi }^{r} x^{r} \right)^{2} \phi _{0}
\end{equation}
and treating the operator$\hat{\chi }$ as an ordinary algebraic quantity, we can \textbf{\textit{formally}} sum up the series on the rhs of equation \eqref{GrindEQ__4_} and recover the Lorentzian function

\begin{equation}
\label{GrindEQ__5_}
\cos _{m} (x)=\dfrac{1}{1+\left({}_{m} \hat{\chi }\, x\right)^{2} } \phi _{0}
\end{equation}
which is assumed to be the UI of $\cos _{m} (x)$.

The possibility of finding a link between rational functions of type \eqref{GrindEQ__1_} and cos-like series is not new and an analogous relationship has been found in a recent past as a part of a study regarding the abstract properties of Gaussian functions \cite{dattoli2023umbral}.

In order to underscore the usefulness of the umbral restyling of nearly cosine functions, we point out that the evaluation of the integrals

\begin{equation}
\label{GrindEQ__6_}
\begin{array}{l}
{I_{m} =\displaystyle \int _{-\infty }^{+\infty }\cos _{m} (x)dx } \\ \\
{m>0}
\end{array}
\end{equation}
can be achieved in a fairly direct way. The use of the same assumption leading to equation \eqref{GrindEQ__5_}, namely $\hat{\chi }$ordinary constant, we find

\begin{equation}
\label{GrindEQ__7_}
\begin{array}{l}
{I_{m} =\displaystyle \int _{-\infty }^{+\infty }\dfrac{1}{1+\left({}_{m} \hat{\chi }\, x\right)^{2} } dx \phi _{0} =\dfrac{\pi }{{}_{m} \hat{\chi }} \phi _{0} =\dfrac{\pi }{(1+m)_{-1} } =\dfrac{\pi \, m!}{\Gamma (m{\kern 1pt} )} =m\pi } \\ \\
{m>0} \nonumber
\end{array}
\end{equation}
the proof relies upon the elementary identity

\[
\int _{-\infty }^{+\infty }\dfrac{1}{1+a\, x^{2} } dx =\dfrac{\pi }{\sqrt{a} }.
\]

\begin{figure}[!ht]
\centering
\includegraphics{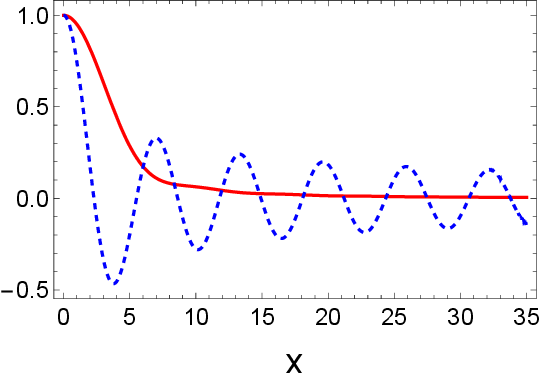}
\caption{$\cos_3(x)$ vs. $x$ (continuous line), $\cos_{1/2}(x)$ (dotted line)}
\label{Fig_1}
\end{figure}

The convergence of the integrals in equation \eqref{GrindEQ__6_} is easily stated, we have checked the validity of equation \eqref{GrindEQ__7_} for two specific examples ($m=1/2,\, 3$ see figure \ref{Fig_1} 1 for the relevant plots).

A further point, we like to touch, concerns the properties, under derivative, of the nearly cosine function \eqref{GrindEQ__1_}. The most natural question is whether a direct link with the associated nearly sine counterpart does hold.

The use of the Laplace transform identity $A^{-1} =\int_{0}^{\infty }e^{-sA} ds$ allows the use of the following integral representation for n. c. functions

\begin{equation}
\label{GrindEQ__8_}
\cos _{m} (x)=\displaystyle \int _{0}^{\infty }e^{-s}  e^{-s\left(\hat{\chi }x\right)^{2} \_ } ds\phi _{0}  \end{equation}
and  keeping the derivative, with respect to x of both sides of equation \eqref{GrindEQ__8_}, we find ($\cos _{m}^{(1)} (x)=\frac{d}{dx} \cos _{m} (x)$)

\begin{equation}
\label{GrindEQ__9_}
\begin{array}{l}
{\cos _{m}^{(1)} (x)=\displaystyle -2x\hat{\chi }^{2} \int _{0}^{\infty }se^{-s}  e^{-s\hat{\chi }^{2} x^{2} \_ } ds\phi _{0} =} \\ \\
{=-2\sum _{r=0}^{\infty }\dfrac{(r+1)!}{r!} \hat{\chi }^{2r+2} x^{2r+1}  } \\ \\
{=-2m!\displaystyle \sum _{r=0}^{\infty }(-1)^{r} \dfrac{(r+1)}{\Gamma (m+2r+3)}  x^{2r+1} }
\end{array}
\end{equation}
which is not directly linked to the n. s. function, if, coherently with the umbral definition in equation \eqref{GrindEQ__4_}, we set

\begin{equation}
\sin _{m} (x)=-x{}_{m} \hat{\chi } \displaystyle \sum _{r=0}^{\infty }(-1)^{r}  \left({}_{m} \hat{\chi }^{r} x^{r} \right)^{2} \phi _{0} =\sum _{r=0}^{\infty }\dfrac{(-1)^{r} }{\left(m+1\right)_{2r+2} }  x^{2r+1} \cos _{m} (x)
\label{GrindEQ__10_}.
\end{equation}
we arrive to the conclusion that

\begin{equation}
\label{GrindEQ__11_}
\cos _{m}^{(1)} (x)\ne -\sin _{m} (x)
\end{equation}
which is by no means surprising and the reasons are simply due to the role, played in the relevant definition, by the umbral operator ${}_{m} \hat{\chi }$.

\begin{figure}[!ht]
         \includegraphics[width=0.48\textwidth]{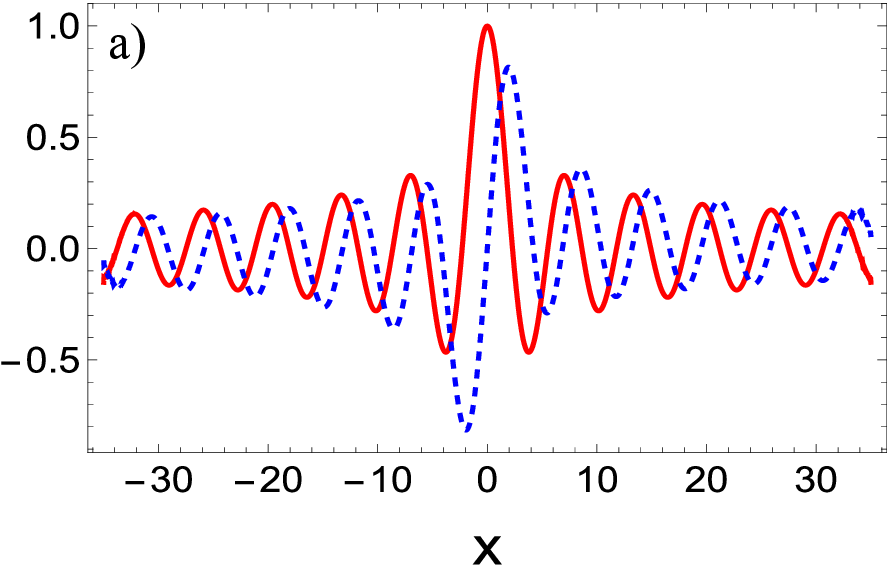}
    \hspace*{\fill}
    \includegraphics[width=0.48\textwidth]{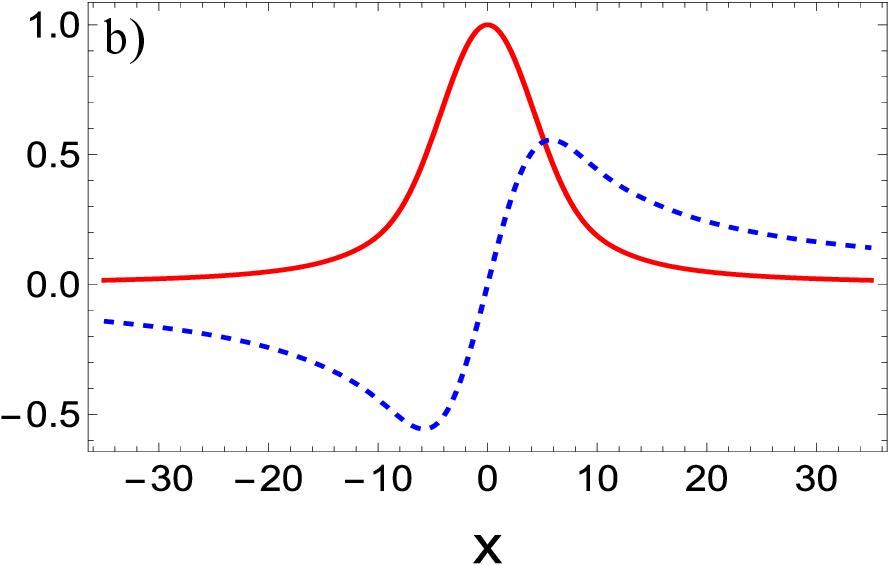}
  \hspace*{\fill} \\
      \includegraphics[width=0.48\textwidth]{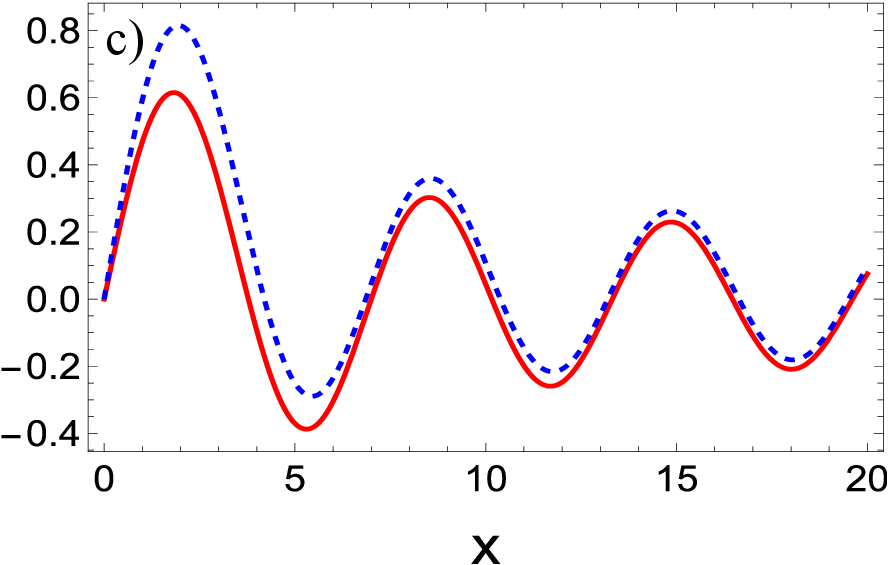}
    \hspace*{\fill}
    \includegraphics[width=0.48\textwidth]{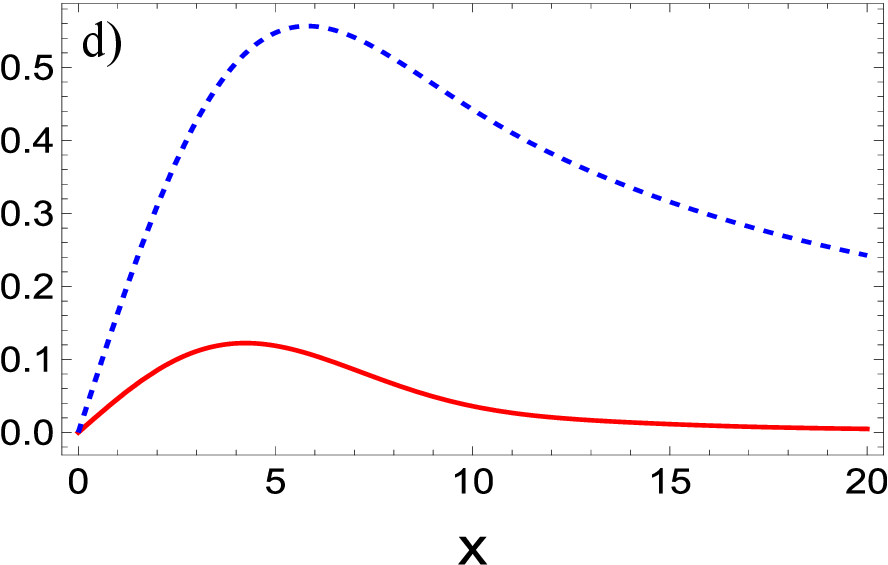}
        \caption{Functions $\cos_m(x)$ (continuous-line), $\sin_m(x)$ (dotted-line)  for different  $m=0.5 (a),5 (b)$. Comparison between $\cos^{(1)}_m(x)$ (continuous-line), $-\frac{d \cos_m(x)}{dx}$ (dotted-line)  for different  $m=0.5 (c),5 (d)$ .}
        \label{Fig_2}
\end{figure}

 In figure \ref{Fig_2} we have reported the behavior of $\cos _{m} (x),\; \sin _{m} (x)$for different values of m and a comparison between $-\frac{d}{dx} \cos _{m} (x),\, \sin _{m} (x)$, which displays an increasing difference for larger m-values.

\begin{figure}[!ht]
         \includegraphics[width=0.32\textwidth]{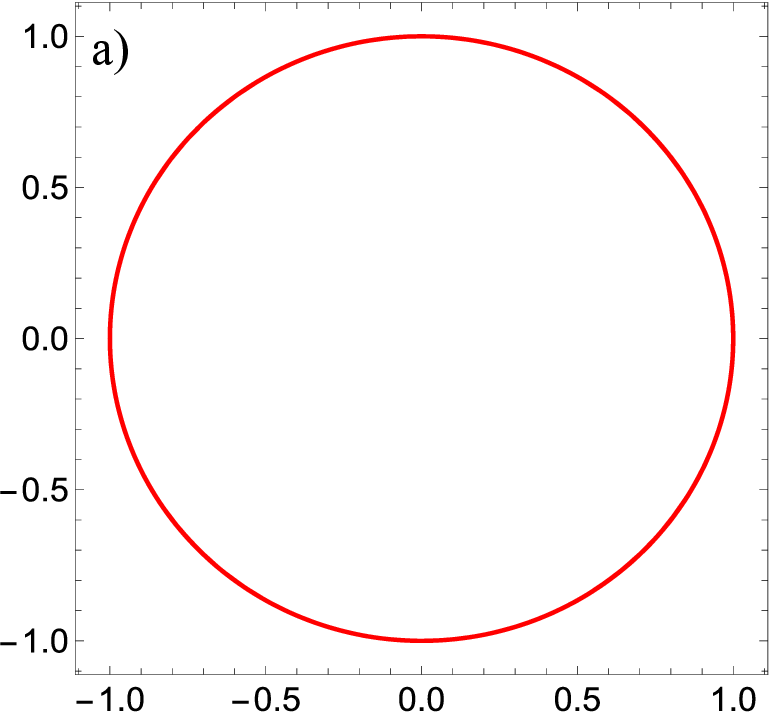}
    \hspace*{\fill}
    \includegraphics[width=0.32\textwidth]{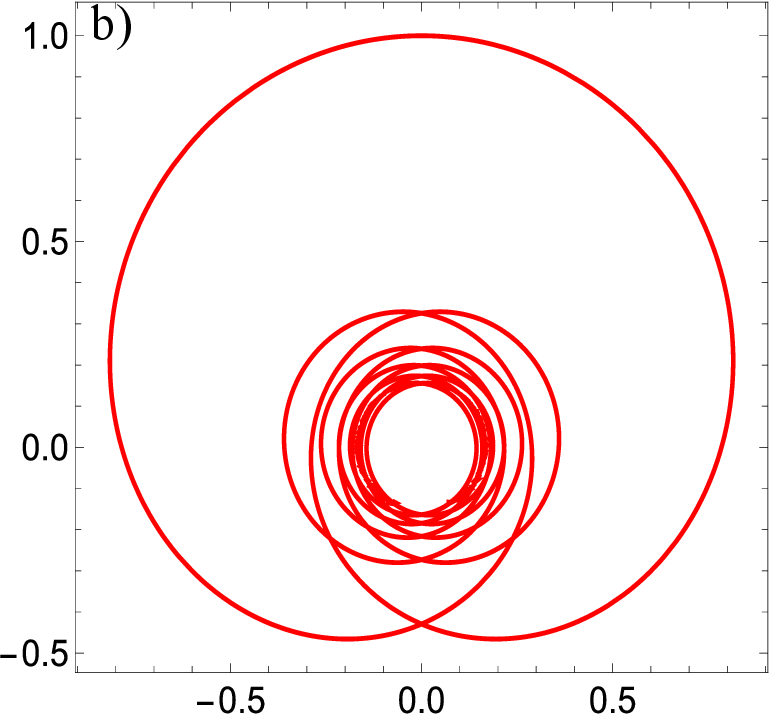}
  \hspace*{\fill}
    \includegraphics[width=0.32\textwidth]{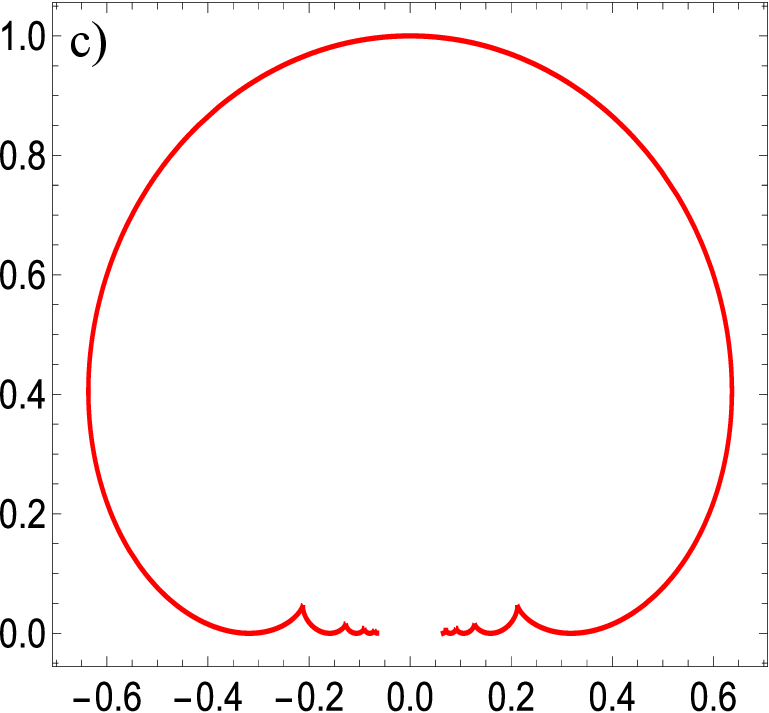}
        \caption{$\cos_m(x)$ vs. $\sin_m(x)$ for different values of $m$: a) $m=0$, b) $m=0.5$, c) $m=2$. }
        \label{Fig_3}
\end{figure}

We like to emphasize that the Nearly Trigonometric Functions (NTF)  functions $\cos _{m} (x),\; \sin _{m} (x)$, are characterized by their own trigonometric circle, as shown in figure \ref{Fig_3}, where we have plotted $\cos _{m} (x)\, vs.\, \sin _{m} (x)$ for different values of m (it should be noted that for m=0 we find the ordinary trigo circumference).

The successive derivatives of the NCF can be comprised in a single formula, which can be obtained by the use of operational methods developed in the past \cite{babusci2019mathematical}. Before proceeding further let us note that the use of the identity

\begin{equation}
\label{GrindEQ__12_}
\begin{array}{l}
{\left(\dfrac{d}{dx} \right)^{m} e^{a\, x^{2} } =H_{m} (2a\, x,a)e^{a\, x^{2} } } \\ \\
{H_{m} (x,y)=n! \displaystyle \sum _{r=0}^{\left\lfloor \frac{n}{2} \right\rfloor }\dfrac{x^{n-2r} y^{r} }{(n-2r)!r!}  }
\end{array}
\end{equation}
we find

\begin{equation}
\label{GrindEQ__13_}
\cos _{m}^{(k)} (x)=(-1)^{k} \displaystyle  \int _{0}^{\infty }H_{k} \left(2xs\hat{\chi }^{2} ,-s\hat{\chi }^{2} \right)\_ e^{-s}  e^{-s\left(\hat{\chi }x\right)^{2} \_ } ds\phi _{0}
\end{equation}
Keeping into account the second of equation \eqref{GrindEQ__12_}, we find

\begin{equation}
\label{GrindEQ__14_}
\begin{array}{l}
{\cos _{m}^{(k)} (x)=(-1)^{k} k!\sum _{l=0}^{\left\lfloor \dfrac{k}{2} \right\rfloor }\dfrac{(-1)^{l} (2x)^{k-2l} }{(k-2l)!l!}  \displaystyle \sum _{r=0}^{\infty }\dfrac{(-1)^{r} }{r!}  \dfrac{x^{2r} m!\Gamma (k-l+r+1)}{\Gamma (1+m+2(r+k)-2l+1)} =} \\ \\
{=(-1)^{k} k!\sum _{l=0}^{\left\lfloor \dfrac{k}{2} \right\rfloor }\dfrac{(-1)^{l} (2x)^{k-2l} }{(k-2l)!l!}  \sum _{r=0}^{\infty }\dfrac{(-1)^{r} x^{2r} }{(m+1)_{2r+2(k-l)} (1+r)_{k-l} }  }
\end{array}
\end{equation}
Thus, finally getting

\begin{equation}
\label{GrindEQ__15_}
\begin{array}{l}
{\cos _{m}^{(k)} (x)=(-1)^{k} k!\displaystyle \sum _{l=0}^{\left\lfloor \dfrac{k}{2} \right\rfloor }\dfrac{(-1)^{l} (2x)^{k-2l} }{(k-2l)!l!}  j_{m,k-l} (x)} \\ \\
{j_{m,\alpha } (x)=\displaystyle \sum _{r=0}^{\infty }\dfrac{(-1)^{r} x^{2r} (1+r)_{\alpha }}{(1+m)_{2(r+\alpha )}  }  }
\end{array}
\end{equation}

The Bessel like nature of the function $j_{m,\alpha} (x)$ is evident. Further comments on this last point will be given in the concluding section.

\section{Further elaborations on NTF}

In this section we like to put in evidence that the formalism we have just outlined is flexible enough to allow extension of the range of functions we are exploring. In particular we note that the function

\begin{equation}
\label{GrindEQ__16_}
\begin{array}{l}
{os_{m}^{(\nu )} (x)=\dfrac{1}{\left[1+\left({}_{m} \hat{\chi }\, x\right)^{2} \right]^{\, \nu } } \phi _{0} =\dfrac{1}{\Gamma (\nu )} \displaystyle \int _{0}^{\infty }s^{\nu -1}  e^{-s} e^{-s\left({}_{m} \hat{\chi }\, x\right)^{2} } ds\phi _{0} } \\ \\
{\nu \in R}
\end{array}
\end{equation}
is easily checked to be characterized by the Taylor expansion

\begin{equation}
\label{GrindEQ__17_}
os_{m}^{(\nu )} (x)=\displaystyle \sum _{r=0}^{\infty }\dfrac{(-1)^{r} \nu _{r} x^{2r} }{(1+m)_{2r} }   \end{equation}
and that

\begin{equation}
\label{GrindEQ__18_}
\displaystyle \int _{-\infty }^{+\infty }os_{m}^{(\nu )} (x)dx =\dfrac{\sqrt{\pi } {}_{m} \hat{\chi }^{-1} }{\Gamma (\nu )} \displaystyle \int _{0}^{\infty }s^{\nu -\dfrac{3}{2} }  e^{-s} ds=m\dfrac{\Gamma \left(\nu -\dfrac{1}{2} \right)}{\Gamma (\nu )} \sqrt{\pi }
\end{equation}

\begin{figure}[!ht]
\centering
\includegraphics{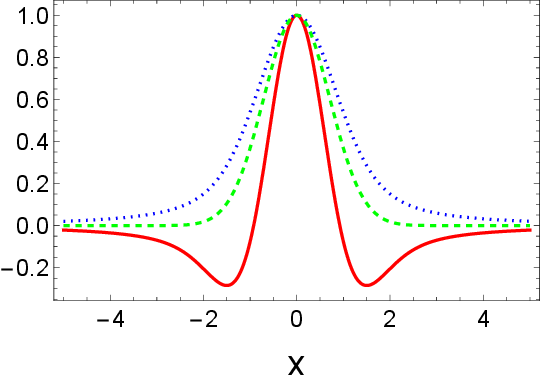}
\caption{The function $e_m(x)$  plotted vs. $x$ for different values of m (continuous line $m=-0.5$, dot $m=0.5$ and dashed $m=0$) }
\label{Fig_4}
\end{figure}

Furthermore, a slight modification of equation \eqref{GrindEQ__5_}, namely

\begin{equation}
\label{GrindEQ__19_}
\begin{array}{l}
{e_{m} (x)=\dfrac{1}{\left[1+{}_{m} \hat{\chi }\, x^{2} \right]} \phi _{0} } \\ \\
{e_{0} (x)=e^{-x^{2} } }
\end{array}
\end{equation}
can be viewed as a kind of Gaussian (see figure \ref{Fig_4}), with longer or smaller tails depending on the specific value of \textit{m}. The utility of this family of functions in statistical problems will be commented elsewhere, as well as their use as complement to NTF.

The UI of $\cos _{m} (x),\, \sin _{m} (x)$, we have chosen in the previous sections, is not compulsory \cite{dattoli2017circular,roman1978umbral,licciardi2022guide,dattoli2023umbral}. Different realizations may be equally useful to look at their properties, from different perspectives and to get further insight into their properties.

An example in this direction is provided by the following exponential realization

\begin{equation}
\label{GrindEQ__20_}
\begin{array}{l}
{\dfrac{\cos _{m} (x)}{m!} =\hat{c}_{-}^{m} e^{-\hat{c}_{-}^{2} \hat{c}_{+} x^{2} } \phi _{+,0} \phi _{-,0} } \\ \\
{\hat{c}_{-}^{s} \phi _{-,0} =\dfrac{1}{\Gamma \left(s+1\right)} ,\, \hat{c}_{+}^{k} \phi _{+,0} =\Gamma (k+1)}
\end{array}
\end{equation}

The proof is easily achieved by expanding the exponential and by applying the operators $\hat{c}_{\pm } $to the respective vacua, namely

\begin{equation}
\label{GrindEQ__21_}
\begin{array}{l}
{\dfrac{\cos _{m} (x)}{m!} =\hat{c}_{-}^{m} \displaystyle \sum _{r=0}^{\infty }\frac{(-1)^{r} c_{-}^{2r} \hat{c}_{+}^{r} }{r!}  x^{2r}\phi _{+,0} \phi _{-,0} =} \\ \\
{=\sum _{r=0}^{\infty }\dfrac{(-1)^{r} c_{-}^{m+2r} \hat{c}_{+}^{r} }{r!}  x^{2r} \phi _{+,0} \phi _{-,0} =\displaystyle \sum _{r=0}^{\infty }\dfrac{(-1)^{r} }{\left(m+2r+1\right)\; !}  x^{2r} }
\end{array}
\end{equation}

The use of the Gauss integral transform

\begin{equation}
\label{GrindEQ__22_}
e^{-b^{2} } =\dfrac{1}{\sqrt{\pi } } \displaystyle \int _{-\infty }^{+\infty }e^{-\xi ^{2} }  e^{2ib\xi x} d\xi  \end{equation}
allows the following umbral restyling of the previous identity

\begin{equation}
\dfrac{\cos _{m} (x)}{m!} =\hat{c}_{-}^{m} \int _{-\infty }^{+\infty }e^{-\xi ^{2} }  e^{-2i\hat{c}_{-} \sqrt{\hat{c}_{+} } \xi \, x} d\xi \phi _{+,0} \phi _{-,0}
\label{GrindEQ__23a_}
\end{equation}
which, in non umbral terms, reads

\begin{equation}
\begin{array}{l}
{\dfrac{\cos _{m} (x)}{m!} =\displaystyle \int _{-\infty }^{+\infty }e^{-\xi ^{2} } e_{m}^{(1/2)} (2ix\xi )\,  d\xi } \\ \\
{e_{m}^{(1/2)} (x)=\displaystyle \sum _{r=0}^{\infty }\dfrac{(-1)^{r} \Gamma \left(\frac{r}{2} +1\right)}{r!\, \Gamma (m+r+1)}  x^{r} }
\end{array}
\label{GrindEQ__23b_}
\end{equation}

\begin{figure}[!ht]
\centering
\includegraphics{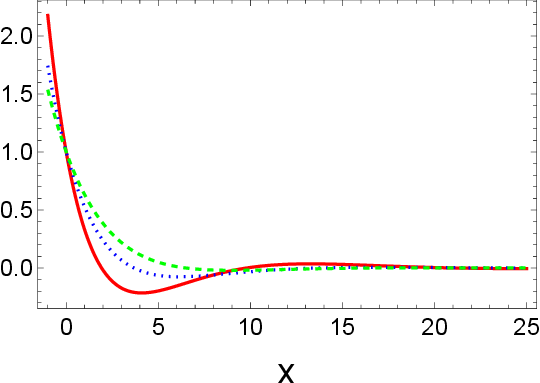}
\caption{Function $e^{(1/2)}_m(-x)$  vs.  $x$ for different values of $m$ (continuous line $m=0$, dot $m=0.5$ and dashed $m=1$).}
\label{Fig_5}
\end{figure}

The behavior of the function \eqref{GrindEQ__23b_}  is provided in figure \eqref{Fig_5}, it resembles the shape of Lennard-Jones-Morse potentials, exploited to model the molecular bonding  \cite{dahl1988morse,royer1968bonding}. The use of nc, ns associated functions to characterize this family of potentials will be the topic of a forthcoming investigation.

\section{Euler NTF and Bessel functions}
\label{sss4}
The functions represented by the series specified in equation \eqref{GrindEQ__1_} can be recognized in terms of ordinary functions and, in correspondence of specific values of the index \textit{m}, we find \cite{sandifer2006euler}

\begin{equation}
\begin{array}{l}
{\cos _{1} (x)=\dfrac{\sin (x)}{x} } \\ \\
{\cos _{2} (x)=2\dfrac{1-\cos (x)}{x^{2} } } \\ \\
{\cos _{3} (x)=6\dfrac{x-\sin (x)}{x^{3} } }
\end{array}
\label{GrindEQ__24a_}
\end{equation}
while, regarding the sine-counterparts, we have checked that

\begin{equation}
\begin{array}{l}
{\sin _{1} (x)=\dfrac{1-\cos (x)}{x} } \\ \\
{\sin _{2} (x)=2\dfrac{x-\sin (x)}{x^{2} } }
\end{array}
\label{GrindEQ__24b_}
\end{equation}
Thus also finding that

\begin{equation}
\label{GrindEQ__25_}
\begin{array}{l}
{\cos _{2} \left(x\right)=\left(\cos _{1} \left(\dfrac{x}{2} \right)\right)^{2} } \\ \\
{\cos _{3} (x)=3\dfrac{\sin _{2} (x)}{x} } \\ \\
{\cos _{2} (x)=2\frac{\sin _{1} (x)}{x} }
\end{array}
\end{equation}

From the previous identities we find, for example
\begin{equation}
\label{GrindEQ__26_}
\begin{array}{l}
{\sin _{2}^{(1)} (x)=\dfrac{\cos _{3} (x)+x\cos _{3}^{(1)} (x)}{3} } \\ \\
{\cos _{2}^{(1)} (x)=\cos _{1} \left(\dfrac{x}{2} \right)\cos _{1}^{(1)} \left(\dfrac{x}{2} \right)}
\end{array}
\end{equation}
which is a partial restatement of the identity reported in equation \eqref{GrindEQ__11_}.

The differential equations satisfied by the nc, ns functions we have studied specific cases as e. g. $\cos _{2,3} (x)$ which are respectively solutions of the following Bessel type ODE

\begin{equation}
\label{GrindEQ__27_}
\begin{array}{l}
{x^{2} z''+4xz'+(2+x^{2} )z-2=0} \\ \\
{x\left[x^{2} z''+6 z'+\left(6 +x^{2} \right)\, z-6\right] =0} \\ \\
{z=\cos _{2,3} (x)}
\end{array}
\end{equation}

The relevant proof is easily achieved by noting e. g. that (see equation \ref{GrindEQ__24b_})

\begin{equation}
\label{GrindEQ__28a_}
\dfrac{2-x^{2} \cos _{2} (x)}{2} =\cos (x)
\end{equation}
Thus finding therefore

\begin{equation}
\label{GrindEQ__29_}
\left(\dfrac{d}{dx} \right)^{2} \left[\dfrac{2-x^{2} \cos _{2} (x)}{2} \right]=-\dfrac{2-x^{2} \cos _{2} (x)}{2}
\end{equation}

The above functions, expressible as combinations of spherical Bessel \cite{andrews1998special,dattoli2017circular} (see below), are almost ubiquitous in classical and quantum optics and, for example, $\cos _{2} (x),-\cos _{2}^{(1)} (x)$ represents the spectral and gain functional dependence of Free Electron devices \cite{colson1990free}. More in general they can be viewed as the absorptive and dispersive parts of scattering processes as further discussed in Appendix I.

 Within this respect an important role is played by the convolution procedures, necessary to include the effect of gain dilution associated with the interplay between the various physical elements constituting the laser device, we have mentioned \cite{colson1990free}.

We denote the convolution of a nearly cosine over a given function $g(x)$, using the standard notation, which for the case of the specific case of a Gaussian reads

\begin{equation}
\label{GrindEQ__28_}
\begin{array}{l}
{\left(\cos _{m} *g\right)\, (x)=\displaystyle \int _{-\infty }^{+\infty }\cos _{m} (\xi ) e^{-\alpha (x-\xi )^{2} } d\xi =} \\ \\
{=\displaystyle \int _{-\infty }^{+\infty }\dfrac{1}{1+\left({}_{m} \hat{\chi }\, \xi \right)^{2} }  e^{-\alpha (x-\xi )^{2} } d\xi \phi _{0} }
\end{array}
\end{equation}
If we split the \textbf{\textit{UI}} as

\begin{equation}
\label{GrindEQ__29a_}
\dfrac{1}{1+\left({}_{m} \hat{\chi }\, \xi \right)^{2} } =\dfrac{1}{2} \left(\dfrac{1}{1+i{}_{m} \hat{\chi }\, \xi } +\dfrac{1}{1-i{}_{m} \hat{\chi }\, \xi } \right)
\end{equation}
We can cast the convolution integral in the form

\begin{equation}
\label{GrindEQ__30_}
\begin{array}{l}
{\displaystyle \int _{-\infty }^{+\infty }\cos _{m} (\xi ) \, e^{-\alpha (x-\xi )^{2} } d\xi =Re \displaystyle \int _{-\infty }^{+\infty }\dfrac{1}{1+i{}_{m} \hat{\chi }\, \xi }  e^{-\alpha (x-\xi )^{2} } d\xi \phi _{0} =} \\ \\
{=\dfrac{e^{-\alpha \, x^{2} } }{2} \sum _{r=0}^{\infty }(-1)^{r} {}_{m} \chi ^{2r}  \displaystyle \int _{-\infty }^{+\infty }\xi ^{2r} e^{-\alpha \xi ^{2} +2\alpha \, x\xi }  d\xi \phi _{0} =} \\ \\
{=e^{-\alpha \, x^{2} } \sum _{r=0}^{\infty }(-1)^{r} {}_{m} \chi ^{2r}  \displaystyle \int _{-\infty }^{+\infty }\xi ^{2r} e^{-\alpha \xi ^{2} +2\alpha \, x\xi }  d\xi \phi _{0} =} \\ \\ {=\displaystyle \sum _{r=0}^{\infty }\dfrac{(-1)^{r} A_{2r} (\alpha ,x)}{(1+m)_{2r} }  ,m>0} \\ \\
{A_{2r} (\alpha ,x)=e^{-\alpha \, x^{2} } \displaystyle \int _{-\infty }^{+\infty }\xi ^{2r} e^{-\alpha \xi ^{2} +2\alpha \, x\xi }  d\xi =e^{-\alpha \, x^{2} } \sqrt{\dfrac{\pi }{\alpha } } \partial _{\beta }^{2r} e^{\frac{\beta ^{2} }{4\, \alpha } } |_{\beta =2\alpha \, x} =} \\ \\
{=\sqrt{\dfrac{\pi }{\alpha } } H_{2r} \left(x,\dfrac{1}{4\alpha } \right)}
\end{array}
\end{equation}

The last result has been obtained by the use of the identity $\partial _{x}^{m} e^{kx^{2} } =H_{m} \left(2kx,k\right)$  (see ref. \cite{babusci2019mathematical} for the necessary details).

We like to underline is the use of equation \eqref{GrindEQ__29a_}, already exploited in ref. \cite{dattoli2023umbral}, where the Umbral theory of Gaussian and Gaussian-like functions has been outlined. The extension of the procedure developed in \cite{dattoli2023umbral} suggests the introduction of the following UI of the exponential function

\begin{equation}
\label{GrindEQ__31_}
\begin{array}{l}
{\exp _{m} \left(ix\right)=\dfrac{1}{1+i{}_{m} \hat{\chi }\, x} \phi _{0} =\displaystyle \sum _{r=0}^{\infty }\left(i{}_{m} \hat{\chi }\, x\right)^{r}  \phi _{0} =} \\ \\
{=\displaystyle \sum _{r=0}^{\infty }\dfrac{\left(i\, x\right)^{r} }{(1+m)_{r} }  }
\end{array}
\end{equation}
which allows the following definitions of the Euler NTF

\begin{equation}
\label{GrindEQ__32_}
\begin{array}{l}
{\cos _{m} \left(x\right)=\dfrac{1}{2} \left(\exp _{m} (ix)+\exp _{m} (-ix)\right)} \\ \\
{\sin _{m} \left(x\right)=\dfrac{1}{2i} \left(\exp _{m} (ix)-\exp _{m} (-ix)\right)}
\end{array}
\end{equation}
consistent with the definitions, given in the previous sections.

It is accordingly evident that

\begin{equation}
\label{GrindEQ__33_}
\cos _{m}^{(k)} =\left(\dfrac{d}{dx} \right)^{k} \exp _{m} \left(ix\right)=Re\left[\sum_{r=0}^{\infty}(-i)^{k+r} \dfrac{\left(k\right)_{r} x^{r} }{(1+m)_{r+k} } \right]
\end{equation}

Significantly simpler than the identity reported in equation \eqref{GrindEQ__15_}, which has been derived to quote that the NTF are linked to Bessel functions in their wider meaning.  Here we substantiate such a statement, we start from the definition in equation \eqref{GrindEQ__1_} and note that, on the basis of the identity [3] we note that

\begin{equation}
x_{(2n)}=2^{2n}\left( \dfrac{x}{2}\right)_n \left( \dfrac{x+1}{2}\right)_n
\end{equation}
we eventually find

\begin{equation}
\cos_m(z)=\displaystyle \sum_{n=0}^{\infty} \dfrac{(1)_n}{\left( \frac{m+1}{2}\right)_n \left( \frac{m+2}{2}\right)_n} \dfrac{(-z^2/4)^n}{n!}
\end{equation}
which can finally cast in the following hyper-geometric form

\begin{equation}
    \cos_m(z)={}_1F_2\left(1;\dfrac{m+1}{2};\dfrac{m+2}{2}; \dfrac{-z^2}{4}\right)
\end{equation}

The tight binding between hypergeometric functions and umbral methods has been discussed in ref. \cite{dattoli2024unveiling}, where the general rules for the transition between  exponential UI and hypergeometrics has been discussed in detail. According to the previous identification, we can get the differential equation of $\cos _{m} \left(x\right)$ using the procedure outlined in \cite{dattoli2024unveiling}. However in order to complete our discussion and substantiate the link with Bessel functions, we note that the Lommel functions, defined in terms of ${}_{1} F_{2} (-)$ reads \cite{gradshteyn1988tables,gray1895treatise}

\begin{equation}
    s_{\mu,\nu}(z)=\dfrac{z^{\mu+1}}{(\mu-\nu+1)(\mu+\nu+1)}{}_{1} F_{2}\left(1;\dfrac{(\mu-\nu+3)}{2},\dfrac{(\mu+\nu+3)}{2};\dfrac{-z^2}{4} \right)
\end{equation}
choosing therefore

\[
\mu=m-\dfrac{3}{2},\;\;\nu=\dfrac{1}{2}
\]
we end up with

\begin{equation}
\label{GrindEQ__38_}
\begin{array}{l}
{\cos _{m} \left(x\right)=m(m-1)x^{\dfrac{1-2m}{2} } s_{m-\frac{3}{2} ,\dfrac{1}{2} } (x)} \\ \\
{m>1} \nonumber
\end{array}
\end{equation}
After using the differential equation for the function $s_{m-\frac{3}{2} ,\frac{1}{2} } (x)$

\begin{equation}
z^2 \dfrac{d^2y}{dz^2}+z \dfrac{dy}{dz}+(z^2-1/4)y=z^{m-1/2}
\end{equation}
yields

\begin{equation}
\label{GrindEQ__40_}
\left[x^{2} \dfrac{d^{2} }{dx^{2} } +2m\, x\dfrac{d}{dx} +(m^{2} -m+x^{2} )\right]\cos _{m} (x)=m(m-1)
\end{equation}
which is a generalization  of the particular cases $m=2,3$ reported in equations \eqref{GrindEQ__27_}.

In this paper we have outlined the embedding of the theory of Euler nearly cosine function within a modern umbral formalism. This effort simplifies the relevant theoretical formulation, their framing within  the more general context of special functions and enhances the novelty of a proposal, dating to more than two centuries ago.

\appendix  \section{Kramers-Kronig relations}

We have already mentioned the relevance of NTF to physical processes involving the scattering processes involving the propagation of electromagnetic waves in dense media and, in general, in spectroscopy, for the modeling of line shapes. A distinctive feature of the complex amplitudes describing these processes are the causality conditions linking their real and imaginary parts, through an Hilbert transform or  Kramers-Kroning relations characterized by the identities reported below \cite{toll1956causality}

\begin{equation}
    \begin{array}{l}
        \sin_m{\omega}= -\dfrac{1}{\pi}\displaystyle Re\left[ \int _{-\infty}^{+ \infty} \dfrac{\cos_m(\omega^{\prime})}{\omega^{\prime}-\omega} d\omega^{\prime} \right]  \\ \\
        \cos_m{\omega}= -\dfrac{1}{\pi}\displaystyle Re\left[ \int _{-\infty}^{+ \infty} \dfrac{\sin_m(\omega^{\prime})}{\omega^{\prime}-\omega} d\omega^{\prime} \right]
    \end{array}
    \label{GrindEQ__40a_}
\end{equation}

\begin{figure}[!ht]
         \includegraphics[width=0.48\textwidth]{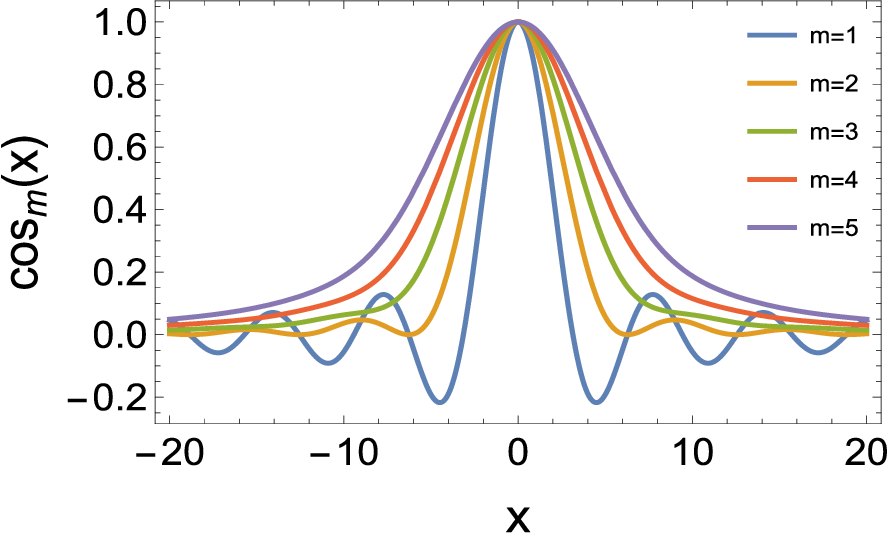}
    \hspace*{\fill}
    \includegraphics[width=0.48\textwidth]{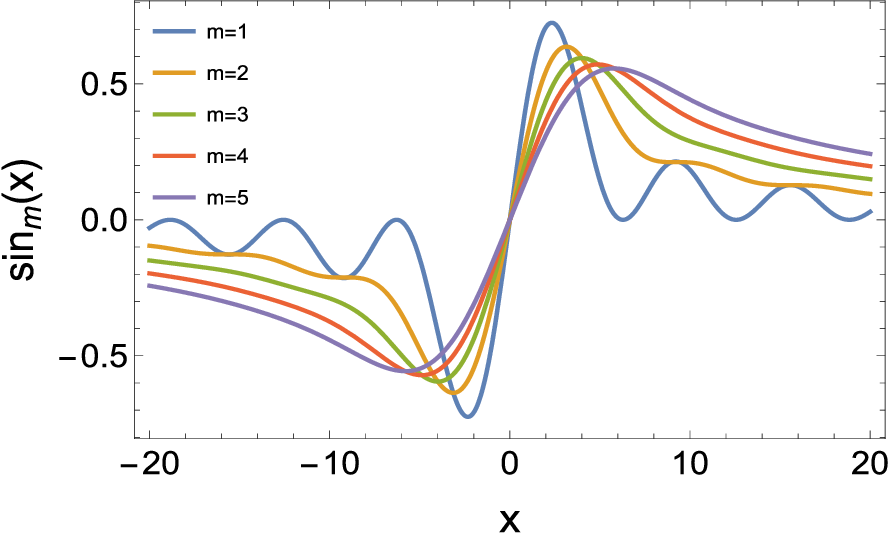}
        \caption{Left: $cos_m(x)$ for different values of $m$; Right: $sin_m(x)$ for different values of $m$.}
        \label{Fig_6}
\end{figure}

A plot of nearly sine and cosine functions is given in figure \eqref{Fig_6}, for different m's.  It seems clear that the nearly cosine functions resemble absorption line shapes, i.e. they could represent the imaginary part of a dielectric susceptibility, while the nearly sine functions resemble the real part of a dielectric susceptibility, related to the real index of refraction of a dense medium.

The proof of the identities \eqref{GrindEQ__40a_} can easily be achieved by noting that
\begin{equation}
\begin{array}{l}
-\dfrac{1}{\pi} \displaystyle \int^{+\infty }_{-\infty }{\dfrac{{cos}_m(\omega^{\prime})}{\omega^{\prime}-\omega}d\omega^{\prime}}=-\dfrac{1}{\pi}\int^{+\infty }_{-\infty }{\dfrac{1}{\omega^{\prime}-\omega}\dfrac{1}{{1+(_m{\hat{\chi}}\omega}^{\prime})^2}\phi_0d\omega^{\prime}}=\\ \\
=\dfrac{1}{i+{}_m{\hat{\chi}}\omega}\phi_0=\dfrac{{}_m{\hat{\chi}}\omega-i}{1+({}_m{\hat{\chi}}\omega)^2}\phi_0
\end{array}
\label{GrindEQ__41_}
\end{equation}

and, eventually realizing that

\begin{equation}
\label{GrindEQ__42_}
Re\left[\dfrac{{}_m\hat{\chi}\omega -i}{1+ ({}_m{\hat{\chi}}\omega )^2} \phi_0 \right]={sin}_m(\omega)
\end{equation}
which demonstrates the first of equations \eqref{GrindEQ__40a_}.  The second equation is easy to prove, starting from the definition at equation \eqref{GrindEQ__10_}.

An analogous result has been obtained in ref. \cite{dattoli2023umbral} where the umbral treatment of the Gaussian functions has naturally led to the link between the ordinary Gaussian and the Dawson function.

The key element extending the HT to these families of functions is a consequence of their UI, realized through a Lorentzian. The functions

\begin{equation}
\label{GrindEQ__43_}
\begin{array}{l}
{l_{s} (x)=\dfrac{1}{1+x^{2} } } \\ \\
{l_{a} (x)=\dfrac{x}{1+x^{2} } }
\end{array}
\end{equation}
provides the most elementary example of Hilbert transform.

\section{Standard Deviation of nearly cosine functions}

Nearly cosine functions are surely normalizable, the latter given by Eq. \eqref{GrindEQ__7_}. However, it is impossible to define a standard deviation for such functions. Indeed, starting from the Lorentzian umbral image at equation \eqref{GrindEQ__5_}, it is straightforward to guess that the integral

\begin{equation}
\label{GrindEQ__36_}
\displaystyle \int^{+\infty }_{-\infty }{\dfrac{{\omega }^{2} }{{\rm 1+(}{{}^{{\rm \ }}_{{\rm m}}{\widehat{\chi }}\omega {\rm )}}^{{\rm 2}}}\phi_0{\rm d}\omega } \end{equation}
does not converge.

Even though sufficient for our purposes the previous statement needs some refinements.

The Lorentzian umbral image has already been exploited to study the umbral properties of Gaussian-like functions. The ordinary Gaussian has indeed been defined as \cite{dattoli2023umbral}

\begin{equation}
\label{GrindEQ__37_}
\begin{array}{l}
{e^{-x^{2} } =\dfrac{1}{1+\hat{c}x^{2} } \phi _{0} } \\ \\ {\hat{c}^{\alpha } \phi _{0} =\dfrac{1}{\Gamma (1+\alpha )} }
\end{array}
\end{equation}
The second order moment can accordingly defined as

\begin{equation}
\label{GrindEQ__38a_}
\displaystyle \int _{-\infty }^{+\infty }x^{2}  e^{-x^{2} } dx=\int _{-\infty }^{+\infty }(\dfrac{x^{2} }{1+\hat{c}x^{2} }  \phi _{0} )dx
\end{equation}
The evaluation of the previous integral goes as it follows

\begin{equation}
\begin{array}{l}
{\displaystyle \int _{-\infty }^{+\infty }(\dfrac{x^{2} }{1+\hat{c}x^{2} }  \phi _{0} )dx=\int _{-\infty }^{+\infty }(\frac{x^{2} }{1+\hat{c}x^{2} }  )dx\phi _{0} =} \\ {=\displaystyle \int _{0}^{+\infty }e^{-s}  \left[\int _{-\infty }^{\infty }x^{2} e^{-s\hat{c}x^{2} }  dx\right]ds\phi _{0} }
\end{array}
\end{equation}

Noting that
\begin{equation}
\left[\displaystyle \int _{-\infty }^{\infty }x^{2} e^{-s\hat{c}x^{2} }  dx\right]\phi _{0} =\dfrac{\sqrt{\pi } }{2} s^{-\dfrac{3}{2} } \hat{c}^{-\dfrac{3}{2} } \phi _{0} =\dfrac{\sqrt{\pi } }{2} \dfrac{s^{-\dfrac{3}{2} } }{\Gamma \left(-\dfrac{1}{2} \right)}
\end{equation}
thus finally getting

\begin{equation}
\label{GrindEQ__39_}
\begin{array}{l}
{\displaystyle \int _{-\infty }^{+\infty }(\frac{x^{2} }{1+\hat{c}x^{2} }  \phi _{0} )dx=\dfrac{\sqrt{\pi } }{2\Gamma \left(-\dfrac{1}{2} \right)} \int _{0}^{+\infty }e^{-s}  s^{-\dfrac{1}{2} -1} ds=} \\ \\
{=\dfrac{\sqrt{\pi } }{2} }
\end{array}
\end{equation}

which seems to be in contradiction with the previous statement. The subtle point is that the derivation of the integral in equation \eqref{GrindEQ__38a_} is based on the following assumptions

\begin{enumerate}
\item  The vacuum can be brought outside the integral

\item  The integration over s and x can be inverted
\end{enumerate}

which holds for forms of the umbral operator allowing the convergence of the integral under study \cite{licciardi2022guide,dattoli2023umbral,babusci2019mathematical}. This last point deserves further investigation and will be discussed with more details elsewhere.

\bibliography{biblio}


\end{document}